\documentclass[12pt]{iopart}
\usepackage{graphicx}
\usepackage{amssymb}
\usepackage[T1]{fontenc}
\usepackage{color}

\begin{document}


\title{Evaluating Gaussianity of heterogeneous fractional Brownian motion}
\author{Micha{\l} Balcerek$^{1,2,*}$, Adrian Pacheco-Pozo$^{1,*}$, Agnieszka Wy{\l}omańska$^2$, and Diego Krapf$^1$}
\address{$^1$ Department of Electrical and Computer Engineering and School of Biomedical Engineering, Colorado State University, Fort Collins, CO 80523, USA.}
\address{$^2$ Faculty of Pure and Applied Mathematics, Hugo Steinhaus Center, Wroc{\l}aw University of Science and Technology, 50-370 Wrocław, Poland.}
\vspace{5pt}
\address{$^*$ These authors contributed equally to this work.}

\ead{michal.balcerek@pwr.edu.pl}


\begin{abstract}
Heterogeneous diffusion processes are prevalent in various fields, including the motion of proteins in living cells, the migratory movement of birds and mammals, and finance. These processes are often characterized by time-varying dynamics, where interactions with the environment evolve, and the system undergoes fluctuations in diffusivity. Moreover, in many complex systems anomalous diffusion is observed, where the mean square displacement (MSD) exhibits non-linear scaling with time. Among the models used to describe this phenomenon, fractional Brownian motion (FBM) is a widely applied stochastic process, particularly for systems exhibiting long-range temporal correlations. Although FBM is characterized by Gaussian increments, heterogeneous processes with FBM-like characteristics may deviate from Gaussianity. In this article, we study the non-Gaussian behavior of switching fractional Brownian motion (SFBM), a model in which the diffusivity of the FBM process varies while temporal correlations are maintained.  To characterize non-Gaussianity, we evaluate the kurtosis, a common tool used to quantify deviations from the normal distribution. We derive exact expressions for the kurtosis of the considered heterogeneous anomalous diffusion process and investigate how it can identify non-Gaussian behavior. We also compare the kurtosis results with those obtained using the Hellinger distance, a classical measure of divergence between probability density functions. Through both analytical and numerical methods, we demonstrate the potential of kurtosis as a metric for detecting non-Gaussianity in heterogeneous anomalous diffusion processes.
\end{abstract}

%
%
\submitto{\JPA}
%
%
%

\section{Introduction \label{sec:intro}}


Heterogeneous diffusion processes are found across diverse fields such as the motion of proteins in living cells \cite{manzo2015weak,akimoto2016universal,weron2017ergodicity,sikora2017elucidating}, the migratory movement of birds and mammals \cite{edelhoff2016path,vilk2022ergodicity},  transport in porous media \cite{berkowitz2002physical}, and finance \cite{alfarano2008time,janczura2013goodness}. In particular, in many complex systems, it is recognized that {environmental heterogeneities cause the dynamics of the tracer to} change over time, where interactions with the environment themselves evolve, the walker visits regions with different conditions, or the molecule of interest changes due to phenomena such as transient binding to a target or dimerization \cite{torreno2016uncovering,munoz2023quantitative,dieball2022scattering}. For systems that undergo Brownian motion, these changes are seen as fluctuations in the diffusivity \cite{jeon2014scaled,miyaguchi2019brownian,hidalgo2021cusp,pacheco2024langevin}. However, complex systems often display anomalous diffusion where the mean square displacement (MSD) exhibits a non-linear scaling with time \cite{metzler2014anomalous,krapf2015mechanisms,norregaard2017manipulation,krapf2019strange,sposini2022towards}. Often the MSD is $\langle r^2(t) \rangle = 2D t^\alpha$, where $\alpha$ is the anomalous exponent and $D$ the generalized diffusion coefficient. 
In such cases, heterogeneous processes can be characterized by changes in the diffusivity or the anomalous exponent. Therefore, one of the first steps in analyzing anomalous diffusion is usually assessing spatio-temporal heterogeneities.

Among the models that describe anomalous diffusion, fractional Brownian motion (FBM) as introduced by Kolmogorov and Mandelbrot is one of the most prevalent stochastic processes \cite{kolmogorov1940wienersche,mandelbrot1968fractional,deng2009ergodic}. It has been used to model the dynamics of tracer particles in mammalian cells \cite{sabri2020elucidating,janczura2021identifying} and other crowded environments \cite{szymanski2009elucidating,weber2010bacterial}, fluorescent molecules deposited on glass \cite{sarfati2020temporally}, ion channels on the surface of neurons \cite{fox2021aging}, telomeres in the cell nucleus \cite{burnecki2012universal,kepten2011ergodicity}, intracellular granules \cite{jeon2011vivo,reverey2015superdiffusion}, and large-scale paths of migrating birds \cite{krapf2019spectral}, to name a few examples. Self-similar anomalous diffusion processes are also found in other diverse systems such as laser cooling \cite{afek2023colloquium} and economic markets \cite{plerou2000economic}. FBM characterizes processes with long-ranged temporal correlations that can exhibit subdiffusion ($\alpha<1$) as well as superdiffusion ($\alpha>1$). Thus, FBM is a very generic stochastic process making it the framework of choice in the analysis of correlated dynamics. Importantly, FBM is a zero-mean Gaussian process and it is, therefore, fully characterized by its covariance function \cite{beran2013long}. In general, it is defined by a Hurst exponent $H$, which is related to the anomalous exponent by $\alpha=2H$.

{In the context of FBM, heterogeneity implies that either $D$ or $H$ change over time (or in space). As a consequence, a process with constant $D$ and $H$, while it has memory, it models dynamics in a homogeneous environment. Within
this context, anomalous diffusion with a
non-linear MSD does not necessarily involve heterogeneity.
Several models have been proposed to consider system heterogeneities. In FBM with random Hurst exponents, the exponent varies from trajectory to trajectory \cite{Balcerek_2022,grzesiek_fbm_fbmre,10.1063/5.0243975}, while in switching FBM (SFBM), the diffusivity or Hurst exponent is a stochastic process  \cite{Balcerek2023,Wang2023,Pacheco2024}.   
Other models arise as the superposition of distinct mechanisms. For example, Brownian yet non-Gaussian diffusion can be modeled using superstatistics, where the diffusion coefficient continuously changes in time \cite{chubynsky2014diffusing, wang2009anomalous, wang2012brownian, chechkin2017brownian,Postnikov2020, pastore2021rapid, Pacheco-Pozo2023}. Other approaches to modeling temporal fluctuations in the tracer dynamics include hybrid models \cite{ribeiro2023interplay} and subordination schemes {\cite{fox2021aging,stanislavsky2014anomalous,chechkin2021relation}}, with the continuous time random walk (CTRW) as a special case {\cite{scher1973stochastic,scher1975anomalous,dybiec2010subordinated}}. A shared aspect of heterogeneous processes is the interplay of different mechanisms arising from changes in the tracer or the  interactions with the environment.} 

Given that the increments of FBM are Gaussian, heterogeneous processes with FBM-like characteristics can be identified by their deviations from Gaussianity \cite{sabri2020elucidating,painter1996evidence,he2016dynamic,lanoiselee2018model,chakraborty2020disorder}. The main tool used for this task is the kurtosis, which for a Gaussian process equals $3$. The kurtosis is often used to classify experimental systems as heterogeneous. However, a characterization of the Gaussianity in heterogeneous FBM is still missing, making it difficult to test the hypothesis that the system under scrutiny is indeed heterogeneous. Furthermore, it would be beneficial to understand whether the kurtosis is sufficient in the quantification of Gaussianity, or if other metrics that directly compare the probability density functions (PDFs) are better suited to the task. It is worth noting that the literature contains numerous robust goodness-of-fit tests for assessing Gaussianity. Prominent examples include the Shapiro-Wilk test~\cite{SWtest}, Jarque-Bera test~\cite{JBtest}, and D'Agostino-Pearson test~\cite{DPtest}. Additionally, a range of methodologies based on evaluating the empirical cumulative distribution function of a random sample have been proposed for testing normality. These include the Kolmogorov-Smirnov test~\cite{KStest}, Cramér-von Mises test~\cite{CMtest}, Kuiper test~\cite{Kuipertest}, Watson test~\cite{Watsontest}, Anderson-Darling test~\cite{ADtest}, and Lilliefors test~\cite{LFtest}. Other approaches have also been proposed~\cite{Iskander,yazici2007comparison,das2016brief,khatun2021applications,razali2011power,thadewald2007jarque}. Although these tests are widely recognized for their effectiveness, the test statistics they employ are often complex, making it challenging to derive the probabilistic properties for the models under analysis. In contrast, kurtosis is a relatively simple statistic and is widely used in various fields to identify impulsiveness (non-Gaussianity) in underlying data. A classical example is condition monitoring, where spectral kurtosis—kurtosis applied to the time-frequency representation of a signal—is the most commonly used technique for signal-based local damage detection~\cite{jerome}.

In this article, we study the Gaussianity of SFBM, where the diffusivity of an underlying FBM process changes while the temporal correlations are maintained throughout the whole realization time. 
We derive exact expressions for the kurtosis of this heterogeneous diffusion process. In particular, we focus on a dichotomous process with random dwell times both for Markovian and non-Markovian switching. Using a combination of analytical and numerical methods, we investigate how kurtosis can reveal Gaussian or non-Gaussian behavior in these systems. We also compare our findings with results obtained from the Hellinger distance~\cite{hellinger1909neue}, a classical divergence measure that quantifies the distance between the PDFs of the considered process and the Gaussian distribution.

The rest of the article is organized as follows. In section \ref{sec1}, we introduce the heterogeneous FBM, and in section \ref{sec3}, we characterize its non-Gaussian behavior using kurtosis. Next, in section \ref{sec4}, we discuss the dichotomous heterogeneous FBM model, referred to as switching FBM, in both Markovian and non-Markovian switching scenarios. In section \ref{s:numerical_study}, we present the results of the numerical simulations. Finally, in section \ref{Hellinger} we discuss the non-Gaussianity of the discussed model in the means of the Hellinger distance, a common measure based on PDF. The final section concludes the article.

\section{Heterogeneous fractional Brownian motion process}\label{sec1}
L\'evy's non-equilibrated integral representation of FBM $B_H(t)$ \cite{Levy1953} is a Gaussian process characterized by a parameter $H \in (0,1)$, known as Hurst exponent, and expressed in term of the Riemann-Liouville integral operator as
\begin{eqnarray}
B_H(t) = \sqrt{4HD} \int_0^t (t-t^{\prime})^{H - 1/2} \, \xi(t^{\prime}) \, dt^{\prime},
\label{eq:levy}
\end{eqnarray}
where $D$ is the generalized diffusion coefficient with units $[\textrm{length}]/ [\textrm{time}^{2H}]$, and $\xi(t)$ is a zero-mean Gaussian white noise with $\delta$-correlations of the form
\begin{eqnarray}
\langle \xi(t_1) \, \xi(t_2) \rangle = \delta(t_1 - t_2).
\label{eq:delta_corr}
\end{eqnarray}
Recently, several modifications of this process have been considered \cite{Balcerek2023,Wang2023,Pacheco2024}. In particular, the case  with the Hurst exponent being a random variable was discussed \cite{Woszczek2024preprint} while the scenario  where the generalized diffusion coefficient is a stochastic process was studied via numerical simulations \cite{Balcerek2023} and analytically \cite{Pacheco2024}.  Along this line, we define FBM with fluctuating diffusivity as a modification of L\'evy's FBM with the generalized diffusion coefficient $D(t)$ being a stochastic process. The process $X(t)$ is written as a Riemann-Liouville fractional integral
\begin{eqnarray}
X(t) = \sqrt{4H} \int_0^t \sqrt{D(t^{\prime})} \, (t-t^{\prime})^{H - 1/2} \, \xi(t^{\prime}) \, dt^{\prime}.
\label{eq:RL_sFBM}
\end{eqnarray}
This process has two sources of randomness, one generated by the Gaussian noise and another by the fluctuations in the diffusion coefficient.

\section{Characterization of non-Gaussianity: Kurtosis}\label{sec3}

The kurtosis is a well-known metric extensively used to assess the Gaussianity of a process or the deviation from it \cite{Meroz2015}. In the literature of diffusion processes, it is defined as the ratio between the fourth moment of the displacements and the square of the second moment (i.e. square of the MSD).
Assuming that the process $X(t)$ initiates at the origin, i.e., $X(t=0) = 0$, we can write the kurtosis as
\begin{eqnarray}
K(t) = \frac{\langle  X^4(t) \rangle}{\langle X^2(t) \rangle^2},
\label{eq:kurtosis_gen}
\end{eqnarray}
which, for a Gaussian process, gives a value of 3.

By evaluating the difference between the kurtosis of a given process and that of a Gaussian one, it is possible to assess (non-)Gaussianity.
The MSD of FBM with fluctuating diffusivities is given by the  convolution \cite{Pacheco2024}
\begin{eqnarray}
\langle X^2(t) \rangle = 4H \int_0^t dt^{\prime} \langle D(t^{\prime}) \rangle \, (t-t^{\prime})^{2H - 1} ,
\end{eqnarray}
where $\langle D(t^{\prime}) \rangle$ is the mean generalized diffusion coefficient. To find the kurtosis, we express the fourth power of the process $X(t)$ as
\begin{eqnarray}
\fl X^4(t) = (4 H )^2 \int_0^t dt_1 \int_0^t dt_2 \int_0^t dt_3 \int_0^t dt_4 \, \sqrt{D(t_1) \, D(t_2) \, D(t_3)\, D(t_4)} \times \nonumber \\
\times \, (t - t_1)^{H - 1/2} \, (t - t_2)^{H - 1/2} \, (t - t_3)^{H - 1/2} \, (t - t_4)^{H - 1/2} \times \nonumber \\
\times \, \xi(t_1) \, \xi(t_2) \,  \xi(t_3) \, \xi(t_4).
\end{eqnarray}
Then, taking the average over the noise we obtain
\begin{eqnarray}
\fl \langle X^4(t) \rangle_{\xi} = (4 H )^2 \int_0^t dt_1 \int_0^t dt_2 \int_0^t dt_3 \int_0^t dt_4  \sqrt{D(t_1) \, D(t_2) \, D(t_3) \, D(t_4)} \times \nonumber \\
\times \, (t - t_1)^{H - 1/2} \, (t - t_2)^{H - 1/2} \, (t - t_3)^{H - 1/2} \, (t - t_4)^{H - 1/2} \times \nonumber \\
\times \, \langle \xi(t_1) \, \xi(t_2) \, \xi(t_3) \, \xi(t_4) \rangle,
\label{eq:noise_avg}
\end{eqnarray}
where we employ the conditional expectation via the nomenclature $\langle X^4(t) \rangle_{\xi}=\langle X^4(t)|D(t) \rangle$.
The last term in the r.h.s. of this expression can be found using Isserlis theorem \cite{Vignat2012} and has the form
\begin{eqnarray}
\fl \langle \xi(t_1) \, \xi(t_2) \, \xi(t_3) \, \xi(t_4) \rangle  =  \langle \xi(t_1) \, \xi(t_2) \rangle \, \langle \xi(t_3) \, \xi(t_4) \rangle + \nonumber \\
+ \langle \xi(t_1) \, \xi(t_3) \rangle \, \langle \xi(t_2) \, \xi(t_4) \rangle + \langle \xi(t_1) \, \xi(t_4) \rangle \, \langle \xi(t_2) \, \xi(t_3) \rangle ,
\end{eqnarray}
which, by means of Eq.~(\ref{eq:delta_corr}), can be rewritten as
\begin{eqnarray}
\fl \langle \xi(t_1) \, \xi(t_2) \, \xi(t_3) \, \xi(t_4) \rangle  =  \delta(t_1 - t_2)  \, \delta(t_3 - t_4) + \nonumber \\
+ \delta(t_1 - t_3) \, \delta(t_2 - t_4) + \delta(t_1 - t_4) \, \delta(t_2 - t_3).
\end{eqnarray}
Plugging this expression in Eq.~(\ref{eq:noise_avg}) we have
\begin{eqnarray}
\fl \langle X^4(t) \rangle_{\xi} = (4 H )^2 \int_0^t dt_1 \int_0^t dt_3  \, D(t_1) \, D(t_3) \, (t - t_1)^{2H - 1} \,(t - t_3)^{2H - 1} + \nonumber \\
+ \, (4 H )^2 \int_0^t dt_1 \int_0^t dt_2  \, D(t_1) \, D(t_2) \, (t - t_1)^{2H - 1} \, (t - t_2)^{2H - 1} + \nonumber \\
+ \, (4 H )^2 \int_0^t dt_1 \int_0^t dt_2 \,  D(t_1) \, D(t_2) \, (t - t_1)^{2H - 1} \, (t - t_2)^{2H - 1}.
\end{eqnarray}
Then, since all variables $t_i$, for $i = 1,2,3$, are essentially dummy variables under the integral sign, we can rewrite this  expression as  
\begin{eqnarray}
\langle X^4(t) \rangle_{\xi} = 3 \, (4 H )^2 \int_0^t dt^{\prime} \int_0^t dt^{\prime\prime} \,  D(t^{\prime} ) \, D(t^{\prime\prime}) \, (t - t^{\prime} )^{2H - 1} \, (t - t^{\prime\prime})^{2H - 1}.
\end{eqnarray}
Next, by taking the average over the diffusion coefficient, one obtains the fourth moment
\begin{eqnarray}
\fl \langle X^4(t) \rangle = \langle \langle X^4(t) \rangle_{\xi} \rangle = 3 \, (4 H )^2 \int_0^t dt^{\prime} \int_0^t dt^{\prime\prime}  \, \langle D(t^{\prime})D(t^{\prime\prime}) \rangle (t - t^{\prime} )^{2H - 1}(t - t^{\prime\prime})^{2H - 1},
\end{eqnarray}
where the iterative expectation is performed first over the noise and then over the fluctuations in the diffusivity.
Finally, the kurtosis (Eq.~(\ref{eq:kurtosis_gen})) takes the form
\begin{eqnarray}
K(t) = 3 \; \frac{\displaystyle  \int_0^t dt^{\prime} \int_0^t dt^{\prime\prime}   \langle D(t^{\prime})D(t^{\prime\prime}) \rangle (t - t^{\prime} )^{2H - 1}(t - t^{\prime\prime})^{2H - 1}}{\displaystyle \left( \int_0^t dt^{\prime} \langle D(t^{\prime}) \rangle (t-t^{\prime})^{2H-1} \right)^2 },
\label{eq:kurtosis_process}
\end{eqnarray}
which depends on the form of the mean generalized diffusion coefficient $\langle D(t^{\prime}) \rangle$ and its covariance function $ \langle D(t^{\prime})D(t^{\prime\prime}) \rangle$.

\section{The dichotomous heterogeneous FBM model: switching FBM}\label{sec4}
{Let us now consider a process where $D(t)$ switches between two states, i.e., a dichotomous process. This heterogeneous model is known as switching fractional Brownian motion (SFBM) with fluctuating diffusivity \cite{Balcerek2023,Pacheco2024}. The two states are characterized by diffusivities $D_{\pm}$ and dwell time distributions $\psi_{\pm}(t)$.} We will consider two types of temporal distributions, one in which all moments of the dwell times exist, and another in which the first moment diverges. To model the first one, we consider an exponential distribution 
\begin{equation}
\psi_{\mathrm{exp}}(t) = \tau^{-1} \exp \left( - t/\tau \right),
\end{equation}
where $\tau$ is the mean dwell time. When both distributions (i.e. $\psi_{\pm}(t)$) are exponential, the dichotomous process is Markovian. To model the distribution lacking the first moment, we consider a power-law distribution that asymptotically behaves as
\begin{eqnarray}
\psi_{\mathrm{PL}}(t) \sim \frac{a}{|\Gamma(-\alpha)|t^{1+\alpha}},
\end{eqnarray}
where $\Gamma(x)$ is the gamma function  \cite{graham1994concrete}, $\alpha \in (0,1)$, and $a$ is a constant with units of $\textrm{s}^\alpha$.

\subsection{Markovian switching}

 We first analyze the case where both dwell times distributions $\psi_{\pm}(t)$ are exponential, i.e., a Markovian process,
\begin{equation}
\psi_{\pm}(t) = \tau_{\pm}^{-1} \exp \left( - t/\tau_{\pm} \right),
\label{eq:exponential}
\end{equation}
where $\tau_{\pm}$ is the mean dwell time of state $\pm$. Given that the process $D(t)$ under these exponential dwell times is Markovian, at long times the initial conditions are forgotten and the process becomes stationary. Thus, the first moment becomes independent of time, and the covariance function $\langle D(t^{\prime})D(t^{\prime\prime}) \rangle$ used in the calculation of kurtosis (Eq.~(\ref{eq:kurtosis_process}))
depends only on the time difference $t^{\prime}-t^{\prime\prime}$. The first moment has the asymptotic form
\begin{eqnarray}
\langle D(t) \rangle = D_+ p_+ + D_- p_-,
\end{eqnarray}
with
\begin{eqnarray}
p_{\pm} = \frac{\tau_{\pm}}{\tau_+ + \tau_-}
\end{eqnarray}
being the probabilities to find a particle in state $\pm$, respectively. The covariance function is then
\begin{eqnarray}
\langle D(t^{\prime})D(t^{\prime\prime}) \rangle = \langle D(t^{\prime}-t^{\prime\prime})D(0) \rangle. 
\end{eqnarray}
Following Ref.~\cite{Miyaguchi2016}, the covariance function can be found in Laplace domain using 
\begin{eqnarray}
\langle D(s)D(0) \rangle = \sum_{h \in \{+,-\}} \sum_{h' \in \{+,-\}} D_{h}D_{h^{\prime}} W_{hh^{\prime}}(s),
\end{eqnarray}
where $W_{hh^{\prime}}(s)$ is the Laplace transform of the transition probability $W_{hh^{\prime}}(t)$, which takes the form
\begin{eqnarray}
W_{\pm \pm}(s) = \frac{p_{\pm}}{s}-\frac{1}{\tau s^2} \frac{[1-\psi_+(s)][1-\psi_-(s)]}{1 - \psi(s)},
\end{eqnarray}
and
\begin{eqnarray}
W_{\pm \mp}(s) = \frac{1}{\tau s^2} \frac{[1-\psi_+(s)][1-\psi_-(s)]}{1 - \psi(s)},
\end{eqnarray}
where $\tau = \tau_+ + \tau_-$ and $\psi(s) = \psi_+(s) \psi_-(s)$. The covariance function in Laplace domain is then expressed as
\begin{eqnarray}
\fl \langle D(s)D(0) \rangle = \frac{p_{+}}{s} D_+^2 + \frac{p_{-}}{s} D_-^2 
- \frac{ p_+ p_-}{ s } (D_+ - D_-)^2 \left(1+\frac{\tau_+\tau_-}{\tau_+ + \tau_-}s \right)^{-1}.
\end{eqnarray}
Finally, going back to time-domain, the covariance function reads
\begin{eqnarray}
 \langle D(t)D(0) \rangle = p_{+} D_+^2  + p_{-} D_-^2  - p_+ p_- (D_+ - D_-)^2 \left[1 - \exp\left( - t/t_c \right) \right],
\label{eq:cov_D_markovian}
\end{eqnarray}
where the correlation time $t_c$ is defined as
\begin{eqnarray}
\label{eq:t_c}
\frac{1}{t_c} = \frac{1}{\tau_+} + \frac{1}{\tau_-}.
\end{eqnarray}

The kurtosis of the dichotomous SFBM with Markovian switching is obtained from Eq.~(\ref{eq:kurtosis_process}). The complete derivation is presented in \ref{sec:AppendixA}, which yields
{\small 
\begin{eqnarray}
 \fl   
 K(t) = 3 \left[   \frac{(D_+^2 p_+ + D_-^2 p_-)}{\left(D_+ p_+ + D_- p_-\right)^2} + 
 \frac{4(D_+ - D_-)^2 p_+ p_- H}{\left(D_+ p_+ + D_- p_-\right)^2} \sum_{n=1}^\infty \frac{(-1)^n }{(4H+n) (2H+1)^{(n)}} \left(\frac{t}{t_c}\right)^n \right],
 \label{eq:kurtosis}
\end{eqnarray}
}\noindent  where $(x)^{(n)} = x (x+1) (x+2) \ldots (x+n-1) = \Gamma(x+n) / \Gamma(x)$ the Pochhammer symbol (also known as falling factorial), and $\Gamma(x)$ is the gamma function.
The series in  Eq. (\ref{eq:kurtosis}) can provide a numerical approximation with any given precision provided a sufficiently high number of elements. Alternatively, one can numerically solve the integral in Eq.~(\ref{eq:kurtosis_process}). A comparison of both approaches and related discussion is presented in section \ref{s:numerical_study}.

\subsection{Non-Markovian switching}
\label{ss:non-markovian}
Next, we consider the case where one or both dwell times distributions $\psi_{\pm}(t)$ have power-law tails with diverging first moment. In contrast to the previous case, the process $D(t)$ is not Markovian. Thus, getting analytical results for such a case is complicated, and we only perform numerical experiments to assess the process. To do so, we consider the Pareto distribution
\begin{equation}
\psi_{PL}(t) = \frac{\alpha t_0^\alpha}{t^{1+\alpha}}, \quad \mathrm{for } t>t_0,
\label{eq:Pareto}
\end{equation}
where $0<\alpha<1$ and $t_0$ is a positive constant with units of time such that $\psi_{PL}(t)=0$ for $t\le t_0$.

\section{Numerical simulations}
\label{s:numerical_study}
\begin{figure}[ht!]
\centering
\includegraphics[width=1\linewidth]{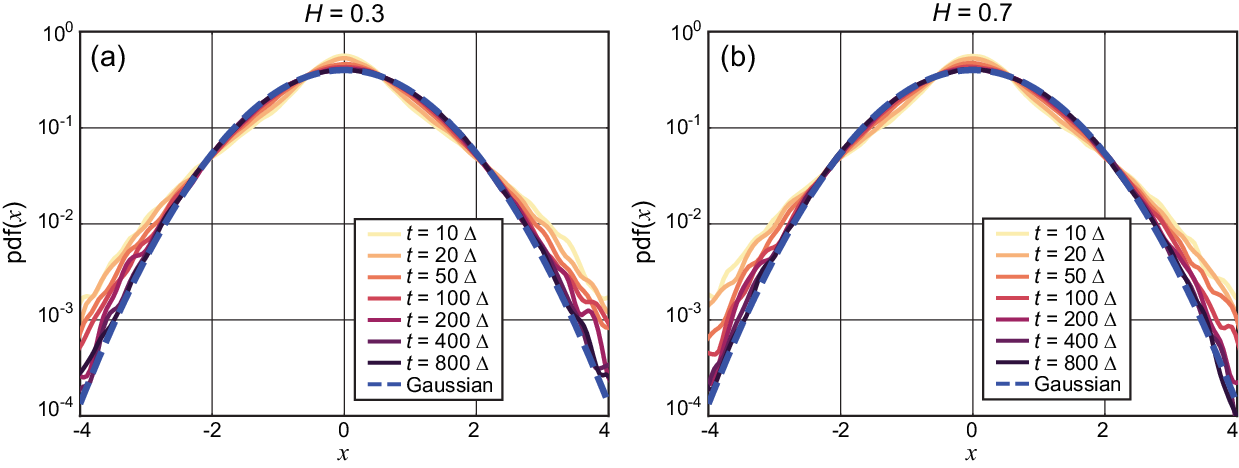}
\caption{Evolution of the PDFs of the normalized position in a dichotomous SFBM with exponential dwell times. Each line corresponds to a specific time, with yellow lines representing shorter times and blue lines representing longer times. The plot PDFs correspond to times equal to $t/\Delta = 10, 20, 50, 100, 200, 400,$ and $800$.  (a) Subdiffusive case with Hurst exponent $H=0.3$. (b) Superdiffusive case with $H=0.7$. On both panels, the dashed line depicts the standard Gaussian PDF.
}
 \label{fig:pdf_ExpExp}
\end{figure}

We simulate switching fractional Brownian motion with fluctuating diffusivities in the same manner as in Ref. \cite{Balcerek2023}. The process is simulated in the interval $[0,T]$, at the specific times $t_i = i \Delta$, with $i = 1,2,\dots, T/\Delta$. We oversample the simulations by dividing each of these time intervals into $M$ subintervals and approximate Eq. (\ref{eq:RL_sFBM}) by using Riemann integration, i.e., as a discrete sum over all the subintervals. The simulations presented in this work employ a time step $\Delta = 10^{-2}$ and oversampling with $M = 4$.
When analyzing data, it is useful to examine the evolution of the position distribution over time, i.e., the propagator of the process. 
Fig. \ref{fig:pdf_ExpExp} shows the PDFs of normalized positions for a subdiffusive ($H=0.3$) and a superdiffusive case ($H=0.7$) measured at different times: $t / \Delta = 10, 20, 50, 100, 200, 400, 800$, where the bright yellow line corresponds to $t / \Delta =10$ and the darkest line corresponds to $t / \Delta =800$. The positions are normalized to their standard deviation so that if they have a Gaussian distribution the normalized positions would have a standard normal distribution as shown by the dashed black line, for comparison. Regardless of the Hurst exponent $H$, in both cases we see a similar rate of convergence toward a Gaussian distribution, where at long times, the distributions of the SFBM process tend to become more Gaussian in shape.

\begin{figure}[ht!]
\centering
\includegraphics[width=1\linewidth]{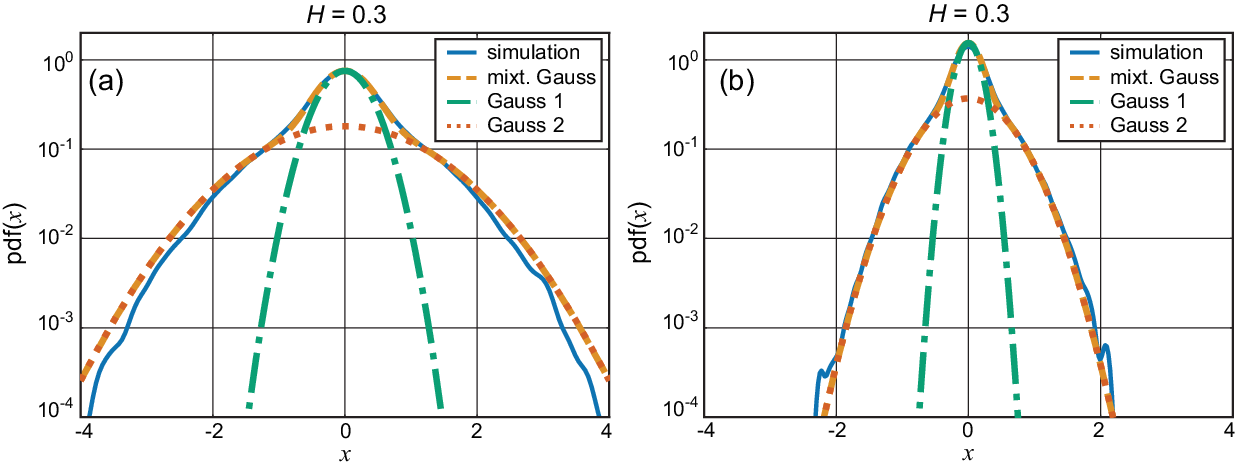}
\caption{PDFs of the position at time $t=3\Delta$ in a dichotomous SFBM with exponential dwell times (blue line). (a) Subdiffusive case with Hurst exponent $H=0.3$. (b) Superdiffusive case with $H=0.7$. On both panels, the dashed yellow line depicts the mixture of Gaussian PDFs described, while dash-dotted green line corresponds to a Gaussian with variance $2D_- t^{2H}$, and dotted orange line represents a Gaussian with variance $2D_+ t^{2H}$. 
}
 \label{fig:pdf_GaussianMixture}
\end{figure}

We visually analyze in some detail the distributions for short times $t$ (before the first switch of the process $D(t)$). The distributions for $H=0.3$ and $H=0.7$ evaluated at $t / \Delta = 3$ are shown in Fig. \ref{fig:pdf_GaussianMixture}. The shapes of the PDFs fit excellently to a mixture of Gaussians \cite{robertson1969some, behboodian1970modes},
\begin{eqnarray*}
    f(x) = p_+ f_+(x) + p_- f_-(x),
\end{eqnarray*}
where the weights $p_+$ and $p_-$ correspond to the probabilities of being in the higher or lower diffusivity states as defined by the initial condition, and $f_+, f_-$ are Gaussian PDFs with zero mean and variances $2D_+t^{2H}$ and $2D_-t^{2H}$, respectively. Specifically, the weights of the presented process are $p_+ = \tau_+/(\tau_+ + \tau_-)$ for the higher diffusivity state and $p_-=1-p_+ = \tau_-/(\tau_+ + \tau_-)$ for the lower diffusivity state. These weights were chosen so that the process $D(t)$ is already equilibrated at $t=0$. 

\begin{figure}[ht!]
\centering
\includegraphics[width=1\linewidth]{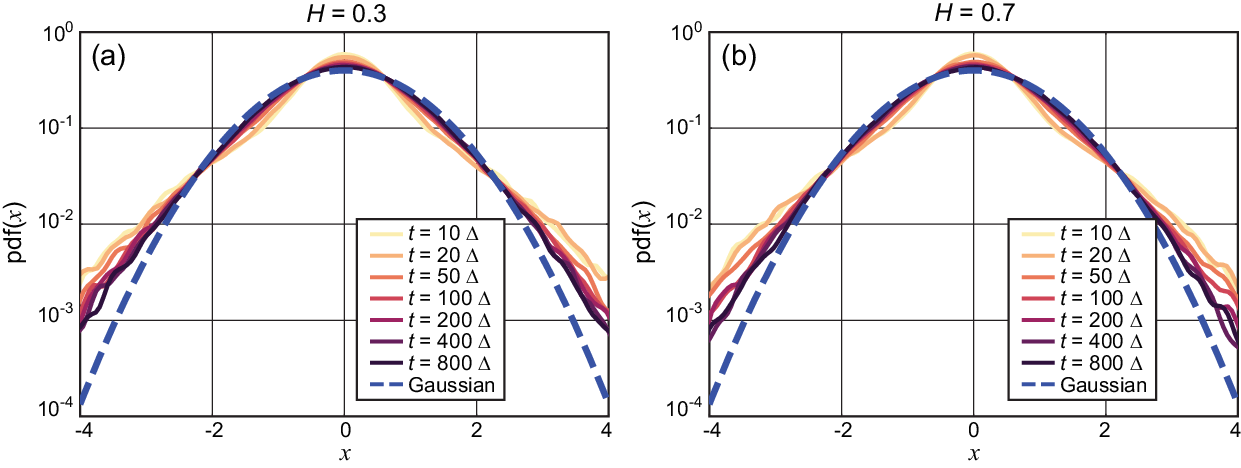}
\caption{Evolution of the PDF of the normalized position in a dichotomous SFBM with one state having a power-law dwell time and the second state, exponential dwell times. Each line corresponds to a specific time, with yellow lines representing shorter times and blue lines representing longer times. The plot PDFs correspond to times equal to $t/\Delta = 10, 20, 50, 100, 200, 400,$ and $800$.  (a) Subdiffusive case with Hurst exponent $H=0.3$. (b) Superdiffusive case with $H=0.7$. On both panels, the dashed line depicts the standard Gaussian PDF. 
}
 \label{fig:pdf_PLExp}
\end{figure}

The situation is substantially different for non-Markovian switching behavior. Fig. \ref{fig:pdf_PLExp} shows the PDF of SFBM where the state with smaller diffusivity ($D_-$) has dwell times drawn from a power-law distribution. The second state still has an exponential distribution. This type of dichotomous process, where one state has a heavy-tailed distribution and the other an exponential distribution, has been studied in multiple physical systems \cite{weigel2013quantifying,sadegh20141,kurilovich2020complex,kurilovich2022non}. In this case,  we do not observe the convergence to a Gaussian distribution. 
Hereafter we refer to the case with one power law and one exponential distribution as PL-Exp and the case with two power laws as PL-PL. Non-Markovian dichotomous models with power-law dwell time distributions in both states are common in biophysics and nanostructures \cite{margolin2004aging,stefani2009beyond,thiel2012anomalous}.

To quantify the Gaussianity of the dichotomous SFBM model, we considered the kurtosis. By examining the kurtosis of the distributions generated by the model, we gain insight into how the shape of the distribution evolves in time. If the distribution becomes more Gaussian, we expect the kurtosis to approach the value of 3, i.e., the kurtosis of a Gaussian distribution. Analyzing the kurtosis can therefore serve as a useful complement to our earlier qualitative analysis of the overall shape of the distribution, as depicted in Figs. \ref{fig:pdf_ExpExp} and \ref{fig:pdf_PLExp}. Together, these two perspectives -- the visual distribution shape and the quantitative kurtosis metric -- can provide a comprehensive understanding of how the statistical properties of the model change over time.

\begin{figure}[ht!]
\centering
\includegraphics[width=0.8\linewidth]{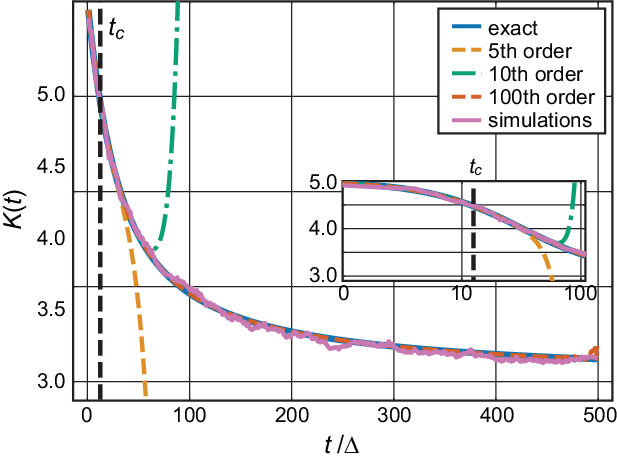}
\caption{Kurtosis for the SFBM with Markovian switching (Exp-Exp) with $H=0.3$. 5th order (dashed yellow line), 10th order (dash-dotted green line), and 100th order (dashed orange line) are shown, where the order describes the number of terms in Eq.~(\ref{eq:kurtosis}). 
The empirical kurtosis (purple line) from 100,000 numerical realizations and the correlation time $t_c$ of the dwell times (vertical dashed black line) are also shown. Inset: Detail of small time. }
 \label{fig:Kapproximation}
\end{figure}

Fig. \ref{fig:Kapproximation} shows the exact kurtosis calculated by numerically integrating Eq.~(\ref{eq:kurtosis_process}) for the Markovian process (Exp-Exp), together with the results obtained by approximating the power series found in Eq.~(\ref{eq:kurtosis}) as a sum of a finite number of elements. A line corresponding to kurtosis estimated from 100,000 trajectories obtained via Monte Carlo simulations is also included.
The figure shows the 5th, 10th, and 100th order approximations for the covariance function $\langle D(t)D(0)\rangle$, where the order corresponds to the number of terms in Eq.~(\ref{eq:kurtosis}).
Importantly, the numerical simulations agree well with the exact derived result, validating our approach to calculate the kurtosis. Furthermore, when 100 terms are used in the series, the approximation matches the exact numerical solution very well within the whole study's range, between $10^{-2}$ and 5 showing that the series indeed converges to the exact result. As fewer terms are used to approximate the sum (Eq.~(\ref{eq:kurtosis})), the approximation can only be used for small $t$. In particular, for $5$ and $10$ terms, the approximation in our example works well up to $t=35\Delta$ and $t =65\Delta$, respectively, with an error smaller than $0.05$. The figure inset zooms on the short time regime ($t \ll t_c$), where $t_c$ is the correlation time as defined in Eq. (\ref{eq:t_c}). 


\begin{figure}[ht!]
\centering
\includegraphics[width=1\linewidth]{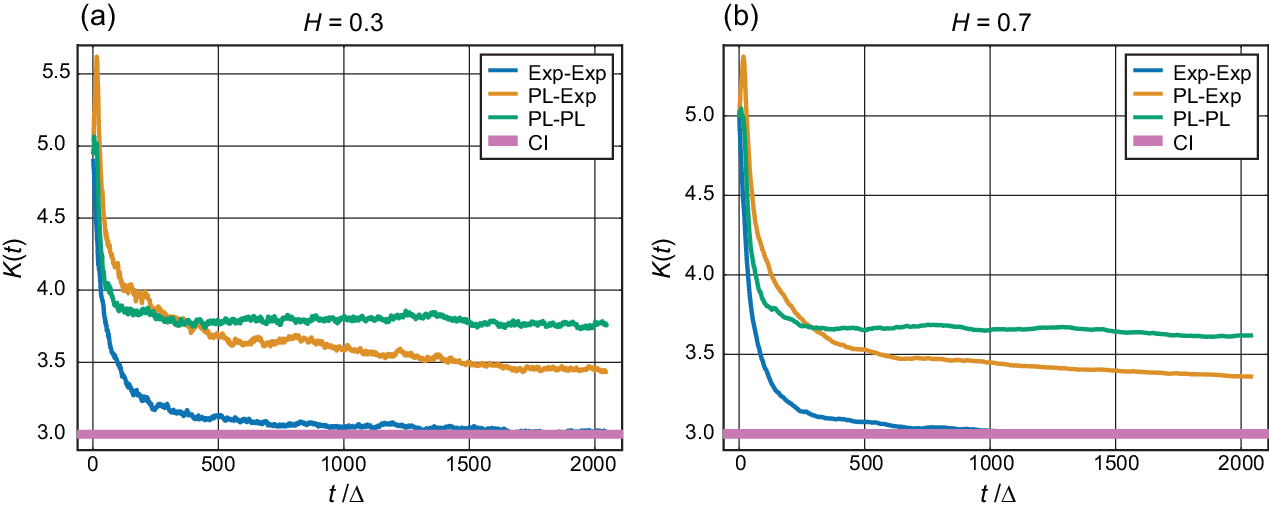}

\caption{Empirical kurtosis for two cases. a) $H=0.3$. b) $H=0.7$. The lines are calculated based on $100 \,000$ trajectories. The shaded area corresponds to the 95\% confidence interval based on 1\,000 samples of length 100\,000.}
 \label{fig:Ksimulation}
\end{figure}

Fig. \ref{fig:Ksimulation} presents the kurtosis $K(t)$ of the SFBM process for both Markovian and non-Markovian state switching as a function of time. We consider different dwell time distributions for the construction of the process $D(t)$.
Specifically, Fig. \ref{fig:Ksimulation}a shows the subdiffusive case with $H=0.3$ and Fig. \ref{fig:Ksimulation}b depicts the superdiffusive case with $H=0.7$. For each of these cases, the following dwell time distributions are considered:
\begin{itemize}
    \item Exp-Exp: Markovian switching, i.e., exponential dwell times in both states.
    \item PL-Exp: Power-law distributed dwell times for the $D_-$ state and exponential dwell times for the $D_+$ state.
    \item PL-PL: Power-law distributed dwell times for both states.
\end{itemize}
The exponential distributions (Eq.~(\ref{eq:exponential})) has an expectation of $\tau_{\pm}=25\Delta$, while the Pareto distributions (Eq.~(\ref{eq:Pareto})) use parameters $\alpha=0.7$ and $\lambda = 15\Delta$. In all cases, the initial condition assumes that each trajectory started in the $D_-$ state with a probability of 0.5, which corresponds to the stationary distribution for the Markovian case.
The purple shaded area in the figures corresponds to the 95\% confidence interval for Gaussian distribution. It is based on 100,000 Gaussian random processes of length $2048$ each, for which the empirical kurtoses were calculated. 
For all the processes with non-Markovian switching, i.e., sub- and super-diffusive as well as PL-Exp and PL-PL distributions, the kurtosis is substantially away from the 95\% confidence interval delineated for Gaussian processes. This shows that the kurtosis can be used to establish the non-homogeneous nature of this family of processes. For both sub- and super-diffusive processes with the PL-Exp dwell times, the kurtosis is observed to decay towards the Gaussian value of 3, but the decay is extremely slow and the kurtosis remains above 3.4 for the whole used realization time ($T=20.48$), that is a trajectory length $T/\Delta=2048$. The situation is more dramatic for PL-PL processes where the kurtosis is observed to converge to a value $3.75$ for $H=0.3$ and to a value $3.60$ for $H=0.7$, for long realization times. In contrast to the non-Markovian switching, the process with Markov switching is clearly non-Gaussian at short times but it becomes Gaussian for long times, i.e., for times $t\gg \tau_{\pm}$. In particular, we observe that the kurtosis falls within the 95\% confidence interval for the Gaussianity test for times $t>13$ ($H=0.3$) and $t>8.7$ (for $H=0.7$).

\begin{figure}[ht!]
\centering
\includegraphics[width=0.8\linewidth]{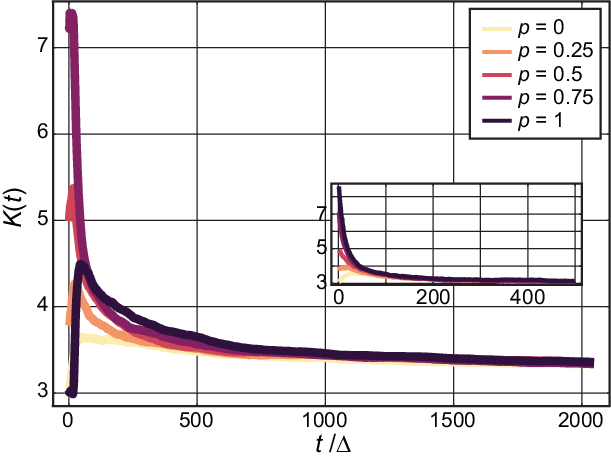}
\caption{Empirical kurtosis for power-law and exponential dwell times (PL-Exp) of SFBM with $H=0.3$. Different lines correspond to different initial conditions ($p$ is the probability of starting in the higher state). The lines are calculated based on $100 \,000$ trajectories. The inset shows the Markovian case (Exp-Exp). Note the smaller time scale in the inset.}
 \label{fig:Ksimulation_wrt_p}
\end{figure}

Fig. \ref{fig:Ksimulation_wrt_p} examines how the evolution of the empirical kurtosis $K(t)$ depends on the initial condition a for PL-Exp SFBM with a subdiffusive Hurst exponent $H = 0.3$. In this case, the low diffusivity state exhibits a power-law waiting time distribution, and the high diffusivity state has an exponential dwell time.
The different lines in the figure correspond to varying initial conditions, where the probability of the process starting in the higher diffusivity state is $p=0, 0.25, 0.5, 0.75,$ and $1$. Since non-stationary distribution is involved in this non-Markovian switching process, the kurtosis values exhibit transient behavior that depends on the initial condition and eventually converges to a master function different from the one expected for the Gaussian distribution (kurtosis $\neq3$).
This behavior highlights how the initial conditions influence the Gaussianity levels at short times. Yet, over longer timescales, the kurtosis dynamics become less sensitive to the initial state, reflecting the dominance of the heavy-tailed waiting time distribution in shaping the overall non-Gaussian statistics of the SFBM process. 
In the Markovian case, as depicted in the inset of Fig.~\ref{fig:Ksimulation_wrt_p}, the empirical kurtosis $K(t)$ converges to 3, in contrast to the non-Markovian scenario. As a result, the kurtosis exhibits minimal sensitivity to initial conditions and rapidly stabilizes to the Gaussian value, reflecting the memoryless nature of the underlying process $D(t)$.

At very short timescales, the kurtosis $K(t)$ can initially increase due to the transient effects introduced by the interplay between the initial conditions and the heterogeneity in the diffusivity process $D(t)$. This effect is further investigated in \ref{appendix:kurtosis_gaussian_mixture}. We show that the kurtosis depends on the probabilities of being in a given state, and the ratio of diffusivities in those states, $c=D_+/D_-$. The dependence of the kurtosis and the diffusivity ratio is particularly pronounced for the non-Markovian case one state having power-law distribution and the second an exponential distribution, as there is no stationary probability distribution for the process $D(t)$. In this scenario, the probability of being in the state with a power-law dwell time approaches 1 at long times. Overall, the kurtosis of the SFBM model is bounded, with a maximum given by
\begin{eqnarray}
    K_{\max} = 3\frac{(c+1)^2}{4c},
\end{eqnarray}
and a minimum value of 3, corresponding to the Gaussian case. 

\begin{figure}[ht!]
\centering
\includegraphics[width=1\linewidth]{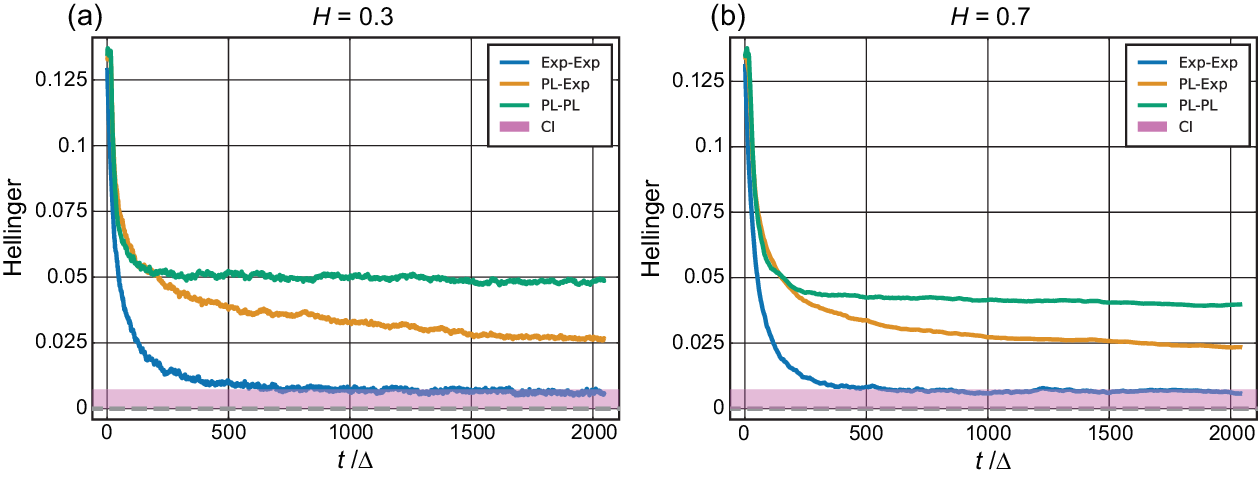}
\caption{Empirical Hellinger distance for two cases of SFBM with different types of dwell times. Panel a) corresponds to $H=0.3$, b) to $H=0.7$. On both panels, the blue lines present the Hellinger distance for the Markovian case, yellow for the dwell times from power-law (lower diffusivity state) and exponential (higher diffusivity state), or power-law for both diffusivity states. The lines are calculated based on $100 \,000$ trajectories, and the pink region corresponds to the 95\% confidence interval.}
 \label{fig:H_distance}
\end{figure}

{
\section{Characterization of non-Gaussianity:  distance between distributions}\label{Hellinger}
In addition to kurtosis we explored alternative metrics for quantifying deviations from Gaussianity.
There are several approaches to measuring the similarity of probability distributions \cite{beran2013long}. 
The Hellinger distance provides a bounded and symmetric measure of divergence between probability distributions. Unlike kurtosis, which focuses solely on the fourth moment, the Hellinger distance evaluates the entire probability density function, thereby it has the potential to capture finer discrepancies between the process under study and a Gaussian reference. This makes it particularly useful in scenarios where higher-order moments or other distributional features significantly deviate from Gaussian behavior.

The square of the Hellinger distance between two distribution with PDFs $p$ and $q$ is defined as \cite{hellinger1909neue, beran1977minimum}
\begin{eqnarray}\label{hellinger}
    H^2(p, q) = 1 - \int_\mathbb{R} \sqrt{p(x)q(x)} dx.
\end{eqnarray}
The measure given in Eq.~(\ref{hellinger}) takes values on interval $[0,1]$, where $0$ corresponds to identical distributions. In general, one advantage of Hellinger distance over other alternatives like Kullback-Leibler divergence \cite{kld} or Bhattacharayya distance \cite{Bd} is that it is both symmetric and bounded, making it more suitable for certain applications where one needs these properties.

In our approach, we replace $p(x)$ by its kernel estimator obtained from the simulated trajectories, choose $q(x)$ as the PDF of a Gaussian distribution with mean $\widehat\mu$ and standard deviation $\widehat\sigma$. These parameters are estimated based on the simulated sample. To check closeness to 0, we also estimate the confidence interval of the Hellinger distance based on a similar approach as in the previous section using the sample of independent identically distributed (iid) random variables from a $\mathcal{N}(0,1)$ distribution.
Fig. \ref{fig:H_distance} shows the Hellinger distance over time $t$ between different cases corresponding to waiting time distributions. Fig. \ref{fig:H_distance}a corresponds to a subdiffusive case with $H=0.3$, and Fig. \ref{fig:H_distance}b to a superdiffusive one with $H=0.7$.
The $y$-axis shows the empirical Hellinger distance values ($0$ to $0.125$), measuring how different these cases of waiting time distributions are from each other as time $t$ evolves from 0 to 20. 
The shaded area corresponds to the 95\% confidence interval for a Gaussian distribution. It was calculated using the same 1\,000 samples of length 100\,000, the same way as the corresponding confidence interval in Fig. \ref{fig:Ksimulation}.  Only the blue line (corresponding to the Markovian case, i.e. exponential dwell times in both states) approaches the confidence interval within the considered time frame. The cases involving a power law dwell time distribution (PL-PL and PL-Exp) do not reach the confidence interval and seem to stabilize very slowly.

Our numerical analysis (Fig. \ref{fig:H_distance}) reveals that the Hellinger distance provides similar results to ones obtained based solely on kurtosis (Fig. \ref{fig:Ksimulation}). Thus, the Hellinger distance does not provide additional information about non-Gaussianity beyond the information obtained evaluating the kurtosis.} 

\


Our analysis was further validated using multiple additional distance metrics. We considered measures based on the cumulative distribution function (Kolmogorov-Smirnov \cite{KStest}, Anderson-Darling \cite{ADtest}, Cram\'er-von Mises \cite{CMtest}) and PDF-based measures (Bhattacharaya distance \cite{Bd}, Kullback-Leibler divergence \cite{kld}). We also performed Kolmogorov-Smirnov goodness-of-fit for Gaussianity \cite{KStest} and considered the related $p$-values. All these methods yielded consistent results similar to the ones obtained with Hellinger distance, reinforcing our conclusions drawn from the kurtosis analysis.

\section{Conclusions}
This study highlights importance of evaluating Gaussianity to understand the heterogeneities in systems exhibiting anomalous diffusion. We investigated the non-Gaussian characteristics of switching fractional Brownian motion, where the diffusivity fluctuates while maintaining temporal correlations. Our focus on switching fractional Brownian motion provides new insights into how fluctuations in diffusivity influence Gaussianity. By deriving exact expressions for kurtosis and validating them with numerical simulations, we demonstrated that kurtosis is a robust and simple metric to quantify deviations from Gaussianity in heterogeneous diffusion processes. Its simplicity is particularly appealing compared to other metrics, as it enables researchers from various domains to effectively characterize complex systems without the need for intricate statistical tests. While kurtosis proved sufficient in capturing the essential aspects of Gaussianity in our analysis, we also explored alternative metrics like the Hellinger distance. These measures provided comparable results but did not offer additional insights, emphasizing kurtosis as a reliable primary choice. 

Our findings indicate that SFBM processes with Markovian switching between states of different diffusivities converge to a Gaussian distribution at long timescales. 
This convergence highlights that, despite the presence of heterogeneities, the memoryless nature of Markovian transitions leads to statistical homogenization over time. Thus, capturing the state-switching dynamics requires sufficient temporal resolution in experimental data. Failure to resolve these transitions can mask underlying heterogeneities, leading to misinterpretations of the system dynamics.
However, this convergence is absent in systems where at least one of the states is characterized by scale-free dynamics, such as power-law dwell time distributions with infinite mean. These systems remain non-Gaussian across all timescales, underscoring the role of heavy-tailed distributions in shaping the statistical properties of diffusion processes. {The non-Gaussianity in processes with scale free dwell times has been discussed extensively for the CTRW with a heavy-tailed distribution of immobilization times. Indeed, a CTRW can be considered as a two-state Brownian process alternating between two different diffusion coefficients in the limit $D_-\to 0$ \cite{e23020231}. For a SFBM, this corresponds to the special case $H=1/2$ and $D_{-}\to 0$. A more general SFBM with $0<H<1$ but $D_{-}\to 0$, can be considered as a combination of the CTRW with FBM, which is usually accomplished via a subordination scheme \cite{krapf2015mechanisms,fox2021aging, dybiec2010subordinated}.}


{The study presented in this article shows that kurtosis is an efficient tool when dealing with SFBM, where the parameter $D$ changes over time. Our results suggest that this statistic should also work well for other heterogeneous anomalous diffusion processes. Previously, other authors have effectively employed kurtosis as a tool for detecting non-Gaussian behavior in different systems ranging from crowded two-dimensional environments \cite{Ghosh_2016} to disordered systems \cite{PhysRevLett.127.120601}. Beyond the model discussed in our article, kurtosis can likely be considered a universal tool useful for heterogeneous models. This will be the subject of further research.}

In conclusion, this work establishes kurtosis as an effective and practical tool for characterizing the non-Gaussian nature of heterogeneous fractional Brownian motion, with implications for broader applications in biophysics, finance, and other complex systems. This study paves the way for future work exploring the  kurtosis {as a universal statistic for non-Gaussianity identification for general heterogeneous processes} and its integration with advanced {classification} methods to automate the characterization of non-Gaussian behaviors.

\appendix

\section{Kurtosis of a Gaussian mixture distribution}
\label{appendix:kurtosis_gaussian_mixture}
Let's consider a random variable $X$ that is a mixture of two zero-mean Gaussian distributions, that is its PDF is given by
\begin{eqnarray*}
    f(x) = pf_1(x) + (1-p)f_2(x),
\end{eqnarray*}
where $f_1$ and $f_2$ correspond to Gaussian PDF with zero means and varianes $\sigma_1^2$ and $\sigma_2^2$, respectively.

To calculate the kurtosis, we first need to consider the second and fourth moment of random variable $X$. Due to properties of mixture distributions we have
\begin{eqnarray}
    \langle X^2 \rangle &= p \sigma_1^2 + (1-p)\sigma_2^2,\\
    \langle X^4 \rangle &= p 3\sigma_1^4 + (1-p)3\sigma_2^4 = 3 (p \sigma_1^4 + (1-p)\sigma_2^4).
\end{eqnarray}
The kurtosis $K = \langle X^4 \rangle/\langle X^2 \rangle^2$ can be thus expressed as
\begin{eqnarray}
    K = 3 \frac{p \sigma_1^4 + (1-p)\sigma_2^4}{(p \sigma_1^2 + (1-p)\sigma_2^2)^2}.
\end{eqnarray}
By introducing the ratio $c = \frac{\sigma_1^2}{\sigma_2^2}$, we can write
\begin{eqnarray}
    \label{eq:kurtosis_wrt_p}
    K = 3 \frac{p c^2 + (1-p)}{(p c + (1-p))^2}=3  \frac{p c^2 + (1-p)}{p^2 c^2 + 2p(1-p)c + (1-p)^2}.
\end{eqnarray}
Considering only the fraction, and substracting denominator from the r.h.s of the above formula we obtain
{\small
\begin{eqnarray*}
    c^2(p-p^2) + (1-p)-(1-p)^2 - 2p(1-p)c &=p(1-p)(c-1)^2 \geq 0. 
\end{eqnarray*}
}
Thus, since the fraction is greater or equal than 1, it follows that the kurtosis $K$ for such a Gaussian mixture is greater or equal to 3, and is equal to 3 only for a single Gaussian distribution ($p=0, p=1$, or $c=1$ which corresponds to the same scales in both Gaussians).

For a fixed $c = \sigma_1^2 / \sigma_2^2$, one can find parameter $p$ that maximises the kurtosis given in Eq. (\ref{eq:kurtosis_wrt_p}). It is equal to
\begin{eqnarray}
    p_{\max} = \frac{1}{c+1} = \frac{\sigma_2^2}{\sigma_1^2+\sigma_2^2}
\end{eqnarray}
for which the maximal kurtosis is equal to
\begin{eqnarray}
    K_{\max} = 3\frac{(c+1)^2}{4c}.
\end{eqnarray}
Different starting levels for kurtosis, presented on Fig. \ref{fig:Ksimulation_wrt_p}, are closely related to this aspect. For the values $\sigma_1^2 = 2D_+, \sigma_2^2 = 2D_-$ considered in the article, $c=10$ and thus $p_{\max}=\frac{10}{11},$ for which the kurtosis is equal to $K = 3\cdot \frac{121}{40} \approx 9$. 

\section{Kurtosis for the Markovian switching}
\label{sec:AppendixA}
For the Markovian case of switching diffusivities, the covariance function is given by Eq. ~(\ref{eq:cov_D_markovian}). By plugging this covariance function into Eq.~(\ref{eq:kurtosis_process}), the fourth moment can be written as
\begin{eqnarray}
\fl \langle X^4(t) \rangle  =  3 (4H)^2 C_1 \int_{0}^{t} \int_{0}^{t}  dt' \, dt'' \, (t - t')^{2H-1} (t - t'')^{2H-1} - \nonumber \\
\fl \qquad - 3 (4H)^2 C_2 \int_{0}^{t} \int_{0}^{t}  dt' \, dt'' \, (t - t')^{2H-1} (t - t'')^{2H-1} \left(1 - \exp\left\{\frac{-|t' - t''|}{t_c}\right\}\right) 
\label{eqn:moment4}
\end{eqnarray}
with
\begin{eqnarray}
C_1 = p_+ \, D_+^2  + p_- \, D_-^2 \qquad \textrm{and} \qquad C_2 = p_+ p_- \, (D_+ - D_-)^2.
\end{eqnarray}
Let us denote the first term in Eq.~(\ref{eqn:moment4}) as $\mathcal{A}$ and the second term as $\mathcal{B}$, and just for simplicity, let us calculate each one of them separately. The first term is straightforward to compute and takes the form
\begin{eqnarray}
\mathcal{A} = 3 (4H)^2 C_1 \left(\int_0^t ds \, (t-s)^{2H-1} \right)^2 = 12 \, C_1 \, t^{4H}.
\label{eq:A}
\end{eqnarray}
The second term can be transformed into the following
\begin{eqnarray}
\label{eq:B4}
\fl \mathcal{B} = 3 (4H)^2 C_2 \, t_c^2 \int_{0}^{t/t_c} \int_{0}^{t/t_c} (t - t_c x)^{2H-1} (t - t_c y)^{2H-1} \left[1 - \exp(-|x-y|)\right] \, dx \, dy,
\end{eqnarray}
by changing to the variables $x = t' / t_c$ and $ y = t'' / t_c$. Then, by defining 
\begin{eqnarray}
\beta = 2H-1 \qquad \textrm{and} \qquad T = t/t_c,
\label{eq:beta_and_T}
\end{eqnarray}
expression (\ref{eq:B4}) can be rewritten as
\begin{eqnarray}
\fl \mathcal{B} = 12 (\beta + 1)^2 C_2 t_c^{2(\beta +1)} \int_0^T dx \int_0^T dy  (T-x)^\beta (T-y)^\beta \left[1 - \exp(-|x-y|)\right].
\end{eqnarray}
Next, we make a final change of variables to $x' = T - x$ and $y' = T - y$ to get
\begin{eqnarray}
\mathcal{B} = 12 (\beta + 1)^2 C_2 t_c^{2(\beta+1)} \int_0^T dx' \int_0^T dy' \, x'^\beta y'^\beta \left[1 - \exp(-|x'-y'|)\right],
\label{eq:B_term}
\end{eqnarray}
which can be rewritten as
\begin{eqnarray}
\mathcal{B} = 12 (\beta + 1)^2 C_2 t_c^{2(\beta+1)} \left[ \int_0^T dx' \int_0^T dy' \, x'^\beta y'^\beta \right. - \nonumber \\
- \left. \int_0^T dx' \int_0^T dy' \, x'^\beta y'^\beta  \exp(-|x'-y'|) \right].
\end{eqnarray}
{
While the first integral inside the square brackets can be solved, the second integral is currently beyond our ability to compute, as the inner integral leads to an incomplete Gamma function multiplied by an exponential. This makes the outer integral impossible to compute directly. However, as we will demonstrate, it is possible to express the solution as a series.} Just for convenience, hereinafter we drop the prime notation in $x'$ and $y'$ and simply write them as $x$ and $y$. Let us then start by considering the Taylor expansion of the exponential term in Eq.~(\ref{eq:B_term}) having the form
\begin{eqnarray}
\exp(-|x-y|) = 1 + \sum_{n=1}^{\infty} \frac{(-1)^{n} |x-y|^n}{n!},
\end{eqnarray}
and plug it back into Eq.~(\ref{eq:B_term}) to get
\begin{eqnarray}
\mathcal{B} = 12 (\beta + 1)^2 C_2 t_c^{2(\beta+1)} \sum_{n=1}^\infty \frac{(-1)^{n+1} \mathcal{A}_n}{n!},
\label{eq:B_term_after}
\end{eqnarray}
where $\mathcal{A}_n$ are defined as 
\begin{eqnarray}
    \label{eq:Ak}
    \mathcal{A}_n \equiv \int_0^T dx \int_0^T dy \, x^\beta y^\beta |x-y|^n,
\end{eqnarray}
for $k=1,2,3, \ldots$. Let us first  concentrate on the $\mathcal{A}_n$ terms and since they involve the integration of an absolute value, let us change the domain of integration from a square to the triangle formed by $x>y$, thus, 
\begin{eqnarray}
\mathcal{A}_n = 2 \int_0^T dx \int_0^x dy \, x^\beta y^\beta (x-y)^n.
\label{eq:A_k}
\end{eqnarray}
Then, let us expand the binomial using the following Newton formula
\begin{eqnarray}
(x-y)^n = \sum_{k=0}^n {n \choose k} (-1)^{n-k} x^k y^{n-k},
\end{eqnarray}
allowing us to write Eq.~(\ref{eq:A_k}) as
\begin{eqnarray}
\mathcal{A}_n = 2 \int_0^T dx \int_0^x dy \, x^\beta y^\beta \sum_{k=0}^n {n \choose k} (-1)^{n-k} x^k  y^{n-k},
\end{eqnarray}
which, after some rearrangement reads
\begin{eqnarray}
\mathcal{A}_n = 2 \sum_{k=0}^n {n \choose k} (-1)^{n-k} \int_0^T dx \, x^{\beta+k} \int_0^x dy \, y^{\beta+n-k}.
\end{eqnarray}
Performing the integral on the $y$ variable leads to 
\begin{eqnarray}
\mathcal{A}_n = 2 \int_0^T dx \, x^{2\beta + n + 1} \sum_{k=0}^n {n \choose k} \frac{(-1)^{n-k} }{\beta+n-k+1},
\end{eqnarray}
where one can clearly see that the integral over $x$ is independent on $k$ and hence can exit the sum. Next, performing the integration over $x$, one gets
\begin{eqnarray}
\mathcal{A}_n = 2 \frac{T^{2\beta+n+2}}{2\beta+n+2} \sum_{k=0}^n {n \choose k}\frac{(-1)^{n-k}}{\beta + n-k + 1}.
\end{eqnarray}
Now, making the change of index $j = n-k$, we can write this last expression as
\begin{eqnarray}
\mathcal{A}_n =2  \frac{T^{2\beta+n+2}}{2\beta+n+2} \sum_{j=0}^n {n \choose j} \frac{(-1)^j}{\beta+1+j}.
\label{eq:Ak_notsimplified}
\end{eqnarray}
Next, we use the fact that
\begin{eqnarray}
\sum_{k=0}^n {n \choose k} \frac{(-1)^k}{\gamma + k} = \frac{n!}{(\gamma)^{(n+1)}},
\end{eqnarray}
with $\gamma$ a positive constant and $(x)^{(n)} = x (x+1) (x+2) \ldots (x+n-1) = \Gamma(x+n) / \Gamma(x) $ the Pochhammer symbol (also known as falling factorial) \cite{graham1994concrete}; to rewrite Eq.~(\ref{eq:Ak_notsimplified}) as
\begin{eqnarray}
\mathcal{A}_n = 2 \frac{n! \, T^{2(\beta+1)+n}}{ [2(\beta+1)+n] (\beta+1)^{(n+1)}}.
\end{eqnarray}
Now, let us replace this result into Eq.~(\ref{eq:B_term}) to get
\begin{eqnarray}
\mathcal{B} = 24 (\beta + 1)^2 C_2 t_c^{2(\beta+1)} \sum_{n=1}^\infty \frac{ (-1)^{n+1}  \, T^{2(\beta+1)+n}}{ [2(\beta+1)+n] (\beta+1)^{(n+1)}},
\end{eqnarray}
which by the definition in Eq.~(\ref{eq:beta_and_T}) takes the form
\begin{eqnarray}
\mathcal{B} = 24 \, (2H)^2 \, C_2 \, t^{4H} \sum_{n=1}^\infty \frac{ (-1)^{n+1}}{ (4H+n) (2H)^{(n+1)}} \left( \frac{t}{t_c} \right)^{n}.
\end{eqnarray}
Then, by joining this result with Eq.~(\ref{eq:A}), the fourth moment reads
\begin{eqnarray}
 \langle X^4(t) \rangle  = 12 t^{4H}  \left[ C_1  -  2 \, (2H)^2 \, C_2 \sum_{n=1}^\infty \frac{ (-1)^{n+1}}{ (4H+n) (2H)^{(n+1)}} \left( \frac{t}{t_c} \right)^{n} \right],
\end{eqnarray}
which can be rewritten as
\begin{eqnarray}
\langle X^4(t) \rangle  = 12 t^{4H}  \left[ C_1  -  4H \, C_2 \sum_{n=1}^\infty \frac{ (-1)^{n+1}}{ (4H+n) (2H+1)^{(n)}} \left( \frac{t}{t_c} \right)^{n} \right],
\end{eqnarray}
by the properties of the Pochhammer symbol.
\\
Finally, after dividing it by the square of the second moment \cite{Pacheco2024}, given by
\begin{eqnarray}
\langle X^2(t) \rangle = 2(D_+ p_+ + D_- p_-) t^{2H} ,
\end{eqnarray}
we obtain the formula for kurtosis given in Eq.~(\ref{eq:kurtosis}) of the main text.

We note, that the series in the numerator is closely related to an integral of the Mittag-Leffler function multiplied by a power-law function. Still, we have decided to leave it as is due to its more complicated form than the one presented in the main text. 

\section*{Acknowledgments}
We thank Prof. Krzysztof Burnecki for many discussions and his valuable insights.\\
Michał Balcerek and Agnieszka Wy\l oma\'nska acknowledge the support from National Science Centre, Poland, via projects No. 2023/07/X/ST1/01139 (MB) and 2020/37/B/HS4/00120 (AW), respectively.  Diego Krapf acknowledges the support from the National Science Foundation Grant 2102832.

\section*{References}

\bibliographystyle{iopart-num}
\bibliography{Manuscript_after_review} 

\end{document}